\documentclass[12pt]{article}
\usepackage{amssymb,amsfonts,amsmath}
\begin{document}
 \hfill\break
\begin{center}\vspace{1.0cm}{\large \bf {Treatment of Landau-Ginzburg Theory with Constraints}}\\

\vspace{1.0cm}{Walaa I.\ Eshraim}\\
{\textit{New York University Abu Dhabi, Saadiyat Island, P.O. Box 129188, Abu Dhabi, U.A.E.}}

\end {center}

\begin{center}
\vspace{1.3cm}{\bf
 Abstract}\end{center} \vspace{0.4cm} 
 Treatment of a singular Lagrangian with constraints using the canonical Hamiltonian approach is studied. We investigate Landau-Ginzburg theory as a constrained system using the Euler-Lagrange equation for the field system and the canonical approach. The equations of motion are obtained as total differential equations in many variables. It is shown that the simultaneous solutions of the Landau-Ginzburg theory with constraints by canonical approach lead to obtaining canonical phase space coordinates and the reduced phase space Hamiltonian without introducing Lagrange multipliers and without any additional gauge fixing condition.
\begin{center}{\bf{keywords}}: Lagrangian and Hamiltonian approach, Singular Lagrangian, Landau-Ginzburg theory.\\
\end{center}

\newpage

\section{Introduction}
$\\$ \indent Singular Lagrangian systems represent a special case
of more general dynamics called constrained systems. A general
feature of constrained systems is the existence of constraints in
their classical configurations.\\
\indent The Lagrangian $L$ of any physical system with $N$ degrees
of freedom is a function of $N$ generalized coordinates $q_i$ and
$N$ generalized velocities $\dot{q}_i$ as well as the time $\tau$,
$$L\equiv(q_{i},\dot{q}_{i},\tau),\quad i=1,...,N.$$ If the
velocities can be expressed in terms of the coordinates and the
momenta, $L$ is referred to as regular, otherwise, it is singular.
Singular Lagrangian systems represent a special case of more
general dynamics called constrained systems. A general feature of
a constrained system is the existence of its classical
configuration.\\
\indent The basic ideas of the classical treatment and
quantization of such systems were initiated and developed by Dirac
\cite{1}. He distinguished between two types of constraints; first- and
second-classes. In the case of unconstrained systems, the
Hamilton-Jacobi theory provides a bridge between classical and
quantum mechanics. The first study of Hamilton-Jacobi equations
for arbitrary first-order actions was carried out by Santilli \cite{2}.
Gitman and Tyutin \cite{3} discussed the canonical quantization of
singular theories as well as the Hamiltonian formalism of gauge
theories in an arbitrary gauge. In the recent past, the canonical method
based on the Hamilton-Jacobi formulation was developed to investigate
singular systems \cite{W1, W3, W4, W5,Eshraim2, W7}. In this formalism, there is no need to
distinguish between first and second constraints as in the Dirac
theory \cite{4,5}. Also, in the canonical method which has been
developed by G\"{u}ler's \cite{6,7}, the equations of motion were
written as total differential equations. In Ref. \cite{NF}, the discrete singular system was treated as
a continuous system. Hamiltonian and Lagrangian formulations are
used together. The
Hamilton-Jacobi formulation of constrained systems has been
studied as seen as in Refs. \cite{H, WH1, WH3}. Moreover, in Refs. \cite{Q1, Q2, Q3, Q4} Hamilton-Jacobi
quantization have been used to obtain the Path integral quantization
for several constraint systems.
 Our aim in this work is to use the Euler-Lagrange
equation to treat the system of a constrained system, the
Landau-Ginzburg theory, and to compare the results to those obtained by Hamilton-Jacobi formulation.\\
\indent The paper is arranged as follows: In section $2$, a brief discussion of the canonical Hamiltonian method is given, together with a treatment of a singular system as a continuous system. Next, in section $3$, Lanfau Ginzburg theory is treated as a singular constrained field system. Finally, in section $4$, several concluding remarks follow.
  
\section{Theoretical framework} $\\$ 
\indent 
In this section, we
review the Hamilton-Jacobi formulation of constrained systems
\cite{1,2}, which the starting point of this method is to consider the
Lagrangian $L\equiv L(q_{i},{\dot{q}}_{i},\tau), i=1,2,\ldots,n$,
with the Hess matrix
\begin{equation}\label{1}
A_{ij}=\frac{\partial^{2}L(q_{i},{\dot{q}}_{i},
\tau)}{{\partial{\dot{q}}_{i}}\>{\partial{\dot{q}}_{j}}}, \qquad
{i,j = 1,2,\ldots,n},
\end{equation}
of rank $(n-r),  r<n$. Then the $r$ momenta are dependent. The
generalized momenta $P_{i}$ corresponding to the generalized
coordinates $q_{i}$ are defined as
\begin{align}
p_{a}&= \frac{\partial{L}}{{\partial{\dot{q}}_{a}}},\qquad {a =
1,2,\ldots,n-r}, \label{2}\\
 p_{\mu}&=\frac{\partial{L}}{{\partial{\dot{q}}_{\mu}}},\qquad {\mu =
n-r+1,\ldots,n},\label{3}
\end{align}
The singularity of the system enables us to solve Eq.(2) for
${\dot{q}}_{a}$ as
\begin{equation}\label{4}
{\dot{q}}_{a}={{\dot{q}}_{a}}(q_{i},{{\dot{q}}_{\mu}},p_{a};\tau)\equiv
\omega_{b}.
\end{equation}
By substituting Eq.(4) into Eq.(3), we obtain the constraints as
\begin{equation}\label{5}
H'_{\mu}= p_{\mu}+H_{\mu}(\tau,q_{i},p_{a})=0,
\end{equation}
where
\begin{equation}\label{6}
H_{\mu}=
-\frac{\partial{L}}{\partial{\dot{q}}_{\mu}}\bigg|_{{\dot{q}}_{a}
\equiv\omega_a}.
\end{equation}

In this formulation the usual Hamiltonian $H_{0}$ is defined as
\begin{equation}\label{7}
H_{0}=-L+ p_{a}{{\dot{\omega}}_a}-{{\dot{q}}_{\mu}}H_{\mu}.
\end{equation}
Like functions $H_{\mu}$, the function $H_{0}$ is not an explicit
function of the velocities ${\dot{q}}_{\nu}$. Therefore, the
Hamilton-Jacobi function $S(\tau,q_{i})$ should satisfy the
following set of Hamilton-Jacobi partial differential equations
(HJPDE) simultaneously for an extremum of the function:
\begin{equation}\label{8}
H'_{\alpha}\bigg(t_{\beta},~q_{\alpha},~P_{i}=\frac{\partial
S}{\partial q_{i}} ,~P_{0}=\frac{\partial S}{\partial
t_{0}}\bigg)=0,
\end{equation}
where\\
$\qquad {\alpha,\beta =0,n-r+1,\ldots,n};\qquad {a =
1,2,\ldots,n-r}$, and
\begin{equation}\label{9}
H'_{\alpha}=p_{\alpha}+H_{\alpha}.
\end{equation}
The canonical equations of motion are given as total differential
equations in variables $t_{\beta}$,
\begin{align}
dq_{p}&=\,\frac{\partial H'_{\alpha}}{\partial p_{p}}\,
dt_{\alpha},\quad {p= 0,1,\ldots,n};\quad {\alpha= 0,n-r+1,\ldots,n},\label{10}\\
dp_{a}&=-\frac{\partial H'_{\alpha}}{\partial q_{a}}\,
dt_{\alpha},\;\qquad a=1,\ldots,n-r,\label{11}\\
dp_{\mu}&=-\frac{\partial H'_{\alpha}}{\partial
q_{\mu}}\,dt_{\alpha}, \,\;\quad \quad \alpha =
0,n-r+1,\ldots,n,\label{12}
\end{align}
\begin{equation}\label{13}
dZ= \left(-H_{\alpha}+p_{a}\frac{\partial H'_{\alpha}}{\partial
p_{\alpha}}\,dt_{\alpha}\right),
\end{equation}
where
\begin{equation}\label{14}
Z\equiv S(t_{\alpha},q_{a}),
\end{equation}
being the action. Thus, the analysis of a constrained system is
reduced to solve equations (10-12) with constraints
\begin{equation}\label{15}
H'_{\alpha}(t_{\beta},~q_{a},~P_{i})=0,\qquad {\alpha,\beta
=0,n-r+1,\ldots,n}.
\end{equation}
Since the equations above are total differential equations,
integrability conditions should be checked. These equations of
motion are integrable if and only if the variations of
$H'_{\alpha}$ vanish identically, that is
\begin{equation}\label{16}
dH'_{\alpha}=0.
\end{equation}
If they do not vanish identically, then we consider them as new
constraints. This procedure is repeated until a complete system is
obtained.\\
 \indent In Ref. [12] the singualr Lagrangian systems are
treated as continuous systems. The Euler-Lagrange equation of a
singular-Lagrangian system is given as
\begin{equation}\label{17}
\frac{\partial }{\partial x_{\alpha}}\bigg[\frac{\partial
L'}{\partial (\partial_{\alpha}q_{\alpha})}\bigg]-\frac{\partial
L'}{\partial
q_{\alpha}}=0,\qquad\partial_{\alpha}q_{\alpha}=\frac{\partial
q_{\alpha}}{\partial x_{\alpha}},
\end{equation}
with constraints
\begin{equation}\label{18}
dG_{\alpha}=-\frac{\partial L'}{\partial x_{\alpha}}\,dt,
\end{equation}
where $L'$ is the "modified Lagrangian" defined as
\begin{equation}\label{19}
L'(x_{\mu},\partial_{\mu}q_{a},\dot{x}_{\mu},q_{a})\equiv
L(x_{\mu},q_{a},\dot{q}_{a}=(\partial_{\mu} q_{a})\dot{x}_{\mu});
\end{equation}
and
\begin{equation}\label{20}
G_{\alpha}=H_{\alpha}\bigg(x_{\mu},q_{a},p_{a}=\frac{\partial
L}{\partial q_{a}}\bigg).
\end{equation}
the solution of Eq. (17), together with the constraints equations
(18), gives us the solution of the system.

\section{The Landau-Ginzburg theory}
\indent The Landau-Ginzburg theory gives an effective
description of phenomenon precisely coincides with scalar quantum
electrodynamics, which is described by the Lagrangian
\begin{equation}\label{21}
\mathcal{L}=-\frac{1}{4}\>F_{\mu\nu} F^{\mu\nu}+
(D_{\mu}\varphi)^\ast D^{\mu}\varphi-k \varphi^{\ast}
\varphi-\frac{1}{4}\lambda (\varphi^{\ast} \varphi)^2,
\end{equation}
where the covariant is given by
\begin{equation}\label{22}
D_{\mu}\varphi=\partial_{\mu}\varphi-ieA_{\mu}\varphi.
\end{equation}
and the electromagnetic tensor is defined as $F^{\mu\nu}=\partial^{\mu} A^{\nu}-\partial^\nu A^\mu$ with the gauge field $A^\nu$. 
In the landau-Ginzburg theory $\varphi$ describes the cooper
pairs. In usual quantum electrodynamics, we would put $k=m^2$,
where $m$ is the effective mass of electron.\\

\subsection {Hamilton-Jacobi formulation of the Landau-Ginzburg theory} 

\indent The Lagrangian function (21) is singular, since
the rank of the Hessian matrix
\begin{equation}\label{23}
A_{ij}=\dfrac{\partial^{2}L}{\partial{\dot{q}}_{i}
\partial{\dot{q}}_{j}}\,,
\end{equation}is three. The canonical momenta are defined as
\begin{equation}\label{24}
\pi^{i}=\frac{\partial L}{\partial {\dot{A}}_{i}}=-F^{0i},
\end{equation}
\begin{equation}\label{25}
\pi^{0}= \frac{\partial L}{\partial{\dot{A}}_{0}}=0,
\end{equation}
\begin{equation}\label{26}
p_{\varphi}= \frac{\partial L}{\partial
\dot{\varphi}}=(D_{0}\varphi)^{\ast}=\dot{\varphi}^{\ast}+ieA_{0}\varphi^{\ast},
\end{equation}
\begin{equation}\label{27}
p_{\varphi^{\ast}}= \frac{\partial L}{\partial
{\dot{\varphi}}^{\ast}}=(D_{0}\varphi)=\dot{\varphi}-i\,e\,A_{0}\,\varphi,
\end{equation}
From Eqs. (24), (26) and (27), the velocities
${\dot{A}}_{i},{\dot{\varphi}}^{\ast}$ and $\dot{\varphi}$ can be
expressed in terms of momenta $\pi_{i}, p_{\varphi}$ and
$p_{\varphi^{\ast}}$ respectively as
\begin{equation}\label{28}
{\dot{A}}_{i}=-\pi_{i}-\partial_{i}A_{0},
\end{equation}
\begin{equation}\label{29}
{\dot{\varphi}}^{\ast}=p_{\varphi}-ieA_{0}{\varphi}^{\ast},
\end{equation}

\begin{equation}\label{30}
{\dot{\varphi}}=p_{{\varphi}^{\ast}}+ieA_{0}{\varphi}.
\end{equation}
The canonical Hamiltonian $H_{0}$ is obtained as
\begin{multline}\label{25}
\qquad
H_{0}=\frac{1}{4}\>F^{ij}F_{ij}-\frac{1}{2}\>\pi_{i}\pi^{i}+\pi^{i}\,\partial_{i}A_{0}+p_{{\varphi}^{\ast}}p_{\varphi}+ieA_{0}{\varphi}p_{\varphi}
\\\,\,\qquad\qquad-ieA_{0}{\varphi}^{\ast}p_{{\varphi}^{\ast}}-(D_{i}\varphi)^{\ast}(D^{i}\varphi)
+k{\varphi}^{\ast}\varphi+\frac{1}{4}\lambda({\varphi}^{\ast}\varphi)^2.
\end{multline}\\
Making use of (7) and (8), we find for the set of HJPDE
\begin{equation}\label{26}
H'_{0}=\pi_{4}+H_{0},
\end{equation}
\begin{equation}\label{27}
H'=\pi_{0}+H=\pi_{0}=0,
\end{equation}
Therefor, the total differential equations for the characteristic
(9-11) obtained as
\begin{multline}
\quad \quad \quad \quad\qquad\qquad\quad dA^{i}=\frac{\partial
H'_{0}}{\partial\pi_{i}}\>dt+\frac{\partial
H'}{\partial\pi_{i}}\>dA^0,\\
=-(\pi^{i}+\partial_{i}A_{0})\,dt,\quad \quad \quad \quad \quad
\quad \quad \quad \quad \,\,\,\,\,\label{28}
\end{multline}

\begin{equation}\label{29}
dA^{0}=\frac{\partial H'_{0}}{\partial\pi_{0}}\>dt+\frac{\partial
H'}{\partial\pi_{0}}\>dA^0=dA^0,
\end{equation}
\begin{multline}
\quad \quad \quad \quad\qquad\qquad\quad d\varphi=\frac{\partial
H'_{0}}{\partial p_{\varphi}}\>dt+\frac{\partial
H'}{\partial p_{\varphi}}\>dA^0,\\
=(p_{{\varphi}^{\ast}}+ieA_{0}\varphi)\,dt,\quad \quad \quad \quad
\quad \quad \quad \quad \quad \,\,\,\,\,\label{30}
\end{multline}
\begin{multline}
\quad \quad \quad \quad\qquad\qquad\quad
d\varphi^{\ast}=\frac{\partial H'_{0}}{\partial
p_{\varphi^{\ast}}}\>dt+\frac{\partial
H'}{\partial p_{\varphi^{\ast}}}\>dA^0,\\
=(p_{\varphi}-ieA_{0}\varphi^{\ast})\,dt,\quad \quad \quad \quad
\quad \quad \quad \quad \quad \,\,\label{31}
\end{multline}
\begin{multline}
\quad \quad \quad \quad d\pi^{i}=-\frac{\partial H'_{0}}{\partial
A_{i}}\>dt-\frac{\partial H'}{\partial A_{i}}\>dA^0,\\
=[\partial_{l}F^{li}+ie(\varphi^{\ast}\partial^{i}\varphi+\varphi\,\partial_{i}\varphi^{\ast})+2e^{2}A^{i}\varphi\varphi^{\ast}]\,dt,\quad
\quad \quad \,\,\label{32}
\end{multline}
\begin{multline}
\quad \quad \quad \quad \quad \quad\,\,\,\,
d\pi^{0}=-\frac{\partial H'_{0}}{\partial
A_{0}}\>dt-\frac{\partial H'}{\partial A_{0}}\>dA^0,\\
=[\partial_{i}\pi^{i}+ie\varphi^{\ast}p_{{\varphi}^{\ast}}-ie\varphi\,p_{\varphi}]\,dt,\quad
\quad \quad \quad \quad \quad \quad\, \label{33}
\end{multline}
\begin{multline}
\quad \quad \quad \quad \quad \quad\,\,\,\,
dp_{\varphi}=-\frac{\partial H'_{0}}{\partial
\varphi}\>dt-\frac{\partial H'}{\partial \varphi}\>dA^0,\\
=[(\overrightarrow{D}\cdot\overrightarrow{D}\varphi)^{\ast}-k\varphi^{\ast}-\frac{1}{2}\lambda\varphi{\varphi^{\ast}}^2-ieA_{0}p_{\varphi}]\,dt,
 \,\,\,\,\,\label{34}
\end{multline}
and
\begin{multline}
\quad \quad \quad \quad \quad \quad\,\,\,\,
dp_{\varphi^{\ast}}=-\frac{\partial H'_{0}}{\partial
\varphi^{\ast}}\>dt-\frac{\partial H'}{\partial \varphi^{\ast}}\>dA^0,\\
=[(\overrightarrow{D}\cdot\overrightarrow{D}\varphi)-k\varphi-\frac{1}{2}\lambda\varphi^{\ast}{\varphi^2}+ieA_{0}p_{\varphi^{\ast}}]\,dt.
\quad \label{35}
\end{multline}
\indent The integrability condition $(dH'_{\alpha}=0)$ implies
that the variation of the constraint $H'$ should be identically
zero, that is
\begin{equation}\label{36}
dH'=d\pi_{0}=0,
\end{equation}
which leads to a new constraint
\begin{equation}\label{37}
H''=\partial_{i}\pi^{i}+ie\varphi^{\ast}p_{{\varphi}^{\ast}}-ie\varphi\,p_{\varphi}=0.
\end{equation}
Taking the total differential of $H''$, we have
\begin{equation}\label{38}
dH''=\partial_{i}d\pi^{i}+iep_{{\varphi}^{\ast}}d\varphi^{\ast}+ie\varphi^{\ast}dp_{{\varphi}^{\ast}}-ie\varphi\,dp_{\varphi}-iep_{\varphi}\,d\varphi=0.
\end{equation}

\subsection{Lagrangian formulation of the Landau- Ginzburg theory}
Let us write the above Lagrangian in the form
\begin{multline}\label{19}
\qquad \mathcal{L}=-\frac{1}{4}\bigg(\frac{\partial
A_{\nu}}{\partial x_{\mu}}-\frac{\partial A_{\mu}}{\partial
x_{\nu}}\bigg)\bigg(\frac{\partial A^{\nu}}{\partial
x^{\mu}}-\frac{\partial A^{\mu}}{\partial x^{\nu}}\bigg)\\+
(\partial_{\mu}+ieA_{\mu})\varphi^\ast
(\partial^{\mu}-ieA^{\mu})\varphi-k \varphi^{\ast}
\varphi-\frac{1}{4}\lambda (\varphi^{\ast} \varphi)^2,
\end{multline}\\
The canonical momenta are defined as
\begin{equation}\label{20}
\pi^{\nu}=\frac{\partial \mathcal{L}} {\partial
(\partial_{\mu}A^\nu)}=-F^{\mu\nu},
\end{equation}
\begin{equation}\label{21}
\pi= \frac{\partial \mathcal{L}} {\partial
(\partial_{\mu}\varphi)}=(D^{\mu}\varphi)^{\ast}=\partial^\mu{\varphi}^{\ast}+ieA^{\mu}\varphi^{\ast}=-H_1,
\end{equation}
\begin{equation}\label{22}
\pi^{\ast}= \frac{\partial \mathcal{L}} {\partial
(\partial_{\mu}\varphi^{\ast})}=(D^{\mu}\varphi)=\partial^\mu{\varphi}-ieA^{\mu}\varphi=-H_2,
\end{equation}
The singular Lagrangian in Eq.(18) can be treated as a continuous
system by introducing
\begin{equation}\label{23}
A_\nu=A_\nu(x_\mu,\varphi,\varphi^{\ast}),\,\varphi=\varphi(x_\mu),\,\varphi^{\ast}=\varphi^{\ast}(x_\mu).
\end{equation}
Let us define the four-dimensional derivative of $A_\nu$ as
\begin{equation}\label{24}
\frac{\partial A_{\nu}}{\partial x_{\mu}}\,\equiv\,\frac{d
A_{\nu}}{d x_{\mu}}=\partial_{\mu}A_{\nu}+{\frac{\partial
A_{\nu}}{\partial \varphi}\frac{\partial \varphi}{\partial
x_{\mu}}}+{\frac{\partial A_{\nu}}{\partial
\varphi^{\ast}}}{\frac{\partial \varphi^{\ast}}{\partial x_{\mu}}}
\end{equation}
The modified Lagrangian $\mathcal{L'}$ becomes
\begin{multline}\label{25}
\mathcal{L'}=\frac{1}{4}\bigg[\partial_{\mu}A_{\nu}+\frac{\partial
A_{\nu}}{\partial \varphi}\,\partial_{\mu}\varphi+\frac{\partial
A_{\nu}}{\partial \varphi^{\ast}}\,\partial_{\mu}\varphi^{\ast}
-\partial_{\nu}A_{\mu}-\frac{\partial A_{\mu}}{\partial
\varphi}\,\partial_{\nu}\varphi-\frac{\partial A_{\mu}}{\partial
\varphi^{\ast}}\,\partial_{\nu}\varphi^{\ast}\bigg]\\\times
\bigg[\partial^{\mu}A^{\nu}+\frac{\partial A^{\nu}}{\partial
\varphi}\,\partial^{\mu}\varphi+\frac{\partial A^{\nu}}{\partial
\varphi^{\ast}}\,\partial^{\mu}\varphi^{\ast}
-\partial^{\nu}A^{\mu}-\frac{\partial A^{\mu}}{\partial
\varphi}\,\partial^{\nu}\varphi-\frac{\partial A^{\mu}}{\partial
\varphi^{\ast}}\,\partial^{\nu}\varphi^{\ast}\bigg]
\\+
(\partial_{\mu}+ieA_{\mu})\varphi^\ast
(\partial^{\mu}-ieA^{\mu})\varphi-k \varphi^{\ast}
\varphi-\frac{1}{4}\lambda (\varphi^{\ast} \varphi)^2
\end{multline}
The Euler-Lagrangian equation for the continuous system (13), for
$q_\alpha\equiv x_{\mu},\varphi,\varphi^{\ast}$ and $q_a\equiv
A_\nu$, becomes
\begin{equation}\label{26}
\frac{\partial }{\partial x_{\mu}}\bigg[\frac{\partial
\mathcal{L'}}{\partial
(\partial_{\mu}A_{\nu})}\bigg]+\frac{\partial }{\partial
\varphi}\bigg[\frac{\partial \mathcal{L'}}{\partial
(\frac{\partial A_{\nu}}{\partial \varphi})}\bigg]+\frac{\partial
}{\partial \varphi^{\ast}}\bigg[\frac{\partial
\mathcal{L'}}{\partial (\frac{\partial A_{\nu}}{\partial
\varphi^{\ast}})}\bigg] -\frac{\partial \mathcal{L'}}{\partial
A_{\nu}}=0,
\end{equation}\label{27}
With the modified Lagrangian $\mathcal{L'}$, Eq.(26) takes the
form
\begin{equation}
\partial_{\mu}F^{\mu\nu}+ie(\varphi^{\ast}\partial^{\nu}\varphi-\varphi\,\partial^{\nu}\varphi)+2e^{2}A^{\nu}\varphi^{\ast}\varphi=0.
\end{equation}
Equation (27) is the first set of Euler-Lagrange equations
obtained from the standard Lagrangian formulation; it is
equivalent to the equations of motion obtained from the canonical
method[9]. The second set of Euler-Lagrange equations of the
standard Lagrangian formulation is obtained by using the
constraint equations (14), that is,
\begin{equation}\label{28}
dG_{1}=-\frac{\partial \mathcal{L'}}{\partial \varphi}\,dx_{\mu}.
\end{equation}
$G_{1}$ is obtained from the Hamiltonian formulation, Eq.(5):
$$G_{1}\equiv H_{1}=-(\partial^\mu{\varphi}^{\ast}+ieA^{\mu}\varphi^{\ast});\qquad and \qquad d\varphi^{\ast}=\frac{\partial\varphi^{\ast}}{\partial x_\mu} dx_\mu$$
Thus, Eq. (28) becomes
\begin{equation}\label{29}
(\overrightarrow{D}\cdot\overrightarrow{D}\varphi)^{\ast}-ie(2A^{\mu}\partial_{\mu}\varphi^{\ast}+\varphi^{\ast}\partial_{\mu}A^{\mu})-k\varphi^{\ast}-\frac{1}{2}\,\lambda\varphi{\varphi^{\ast}}^2=0.
\end{equation}
Similarly, from Eq.(5), we have $$G_{2}\equiv
H_{2}=-(\partial^\mu{\varphi}-ieA^{\mu}\varphi)\qquad and \qquad
d\varphi=\frac{\partial\varphi}{\partial x_\mu} dx_\mu$$. Then, by
using Eq.(14), we get
\begin{equation}\label{30}
dG_{2}=-\frac{\partial \mathcal{L'}}{\partial
\varphi^{\ast}}\,dx_{\mu},
\end{equation}
above equation becomes
\begin{equation}\label{31}
(\overrightarrow{D}\cdot\overrightarrow{D}\varphi)+ie(2A^{\mu}\partial_{\mu}\varphi+\varphi\,\partial_{\mu}A^{\mu})-k\varphi-\frac{1}{2}\,\lambda\varphi^{\ast}{\varphi}^2=0.
\end{equation}
Equations (29) and (31) are the second set of Euler-Lagrange
equations of the standard formulation.

\section {Conclusion}
\indent The Lagrangian of Landau-Ginzburg theory gives an
effective description of phenomenon precisely coincides with
scalar quantum electrodynamics. This system is studied as a singular Lagrangian
using the Euler-Lagrange equation and the canonical Hamiltonian approach (Hamilton-Jacobi approach). The system is treated as a
continuous field system with constraints. It is shown that this
treatment is in exact agreement with the general approach. Our
formalism is a mixture of the Hamiltonian and Lagrangian
formulations. In the general approach, the constraint equations
can be obtained from the Euler-Lagrange equations; whereas in the
treatment of singular Lagrangian as fields, the constraints can be
determined from Eq. (14), which is obtained with the help of the
canonical Hamiltonian formalism. The equations of motion are obtained as
partial differential equations, which are equivalent to those
equations obtained from the canonical Hamiltonian approach.

\end{document}